\documentclass[10pt,conference]{IEEEtran}

\usepackage{amssymb}
\usepackage{graphicx}
\usepackage{amsmath}

\begin{document}

\title{Bounds on the Capacity of the Blockwise Noncoherent APSK-AWGN Channels}

\author{\authorblockN{Daniel C. Cunha and Jaime Portugheis}
\authorblockA{Department of Communications,
School of Electrical and Computer Engineering\\
State University of Campinas\\
C.P. 6101, 13083-852, Campinas-SP, Brazil \\
Emails: dcunha@decom.fee.unicamp.br~,~jaime@decom.fee.unicamp.br}}

\maketitle

\begin{abstract}
Capacity of \emph{M-ary Amplitude and Phase-Shift Keying} ($M$-APSK) over an
\emph{Additive White Gaussian Noise} (AWGN) channel that also introduces an unknown
carrier phase rotation is considered. \ The phase remains constant over a block of $L$
symbols and it is independent from block to block. Aiming to design codes with equally
probable symbols, uniformly distributed channel inputs are assumed. Based on results of
Peleg and Shamai for \emph{M-ary Phase Shift Keying} ($M$-PSK) modulation, easily
computable upper and lower bounds on the effective M-APSK capacity are derived. For
moderate $M$ and $L$ and a broad range of \emph{Signal-to-Noise Ratios} (SNR's), the
bounds come close together. As in the case of $M$-PSK modulation, for large $L$ the
coherent capacity is approached.
\end{abstract}

\section{Introduction}\label{sec:intro}
Coherent reception is not possible for many bandpass transmission systems.  In these
systems, it is commonly assumed that the unknown carrier phase rotation is constant over
a block of $L$ symbols and independent from block to block.  One approach adopted to
solve the problem of detection of information transmitted over these systems is
\emph{Multiple Symbol Differential Detection} (MSDD) \cite{Divsalar90}.  The system
modulation is usually \emph{M-ary Phase Shift Keying} ($M$-PSK), but in the case of high
spectral efficiencies, \emph{M-ary Amplitude and Phase-Shift Keying} ($M$-APSK) with
independent phase and amplitude modulations is preferable \cite{Cahn60}, \cite{Lampe99}.

The capacity of a noncoherent AWGN channel in the case of input symbols drawn from an
$M$-PSK modulation has been investigated by Peleg and Shamai \cite{Peleg98}. It was shown
that capacity can be achieved by uniformly, independently and identically distributed
(u.i.i.d) symbols. For the case of detection with an overlapping of one symbol, upper and
lower bounds on capacity were given. Aiming to design codes with equally probable
symbols, u.i.i.d channel inputs are here assumed. In this case, the capacity is
denominated effective. Extending the results for $M$-PSK modulation, easily computable
upper and lower bounds on the effective $M$-APSK capacity are derived.

The outline of the paper is as follows. In Section \ref{sec:apsk}, we compare the
effective capacity of some APSK constellations in the case of coherent reception. In
Section \ref{sec:channel_mod}, we define the noncoherent channel model. Section
\ref{sec:bounds} describes the derivation of the upper and lower bounds on the capacity
of the noncoherent APSK channels. Section \ref{sec:conclusion} concludes the paper
presenting numerical results.

\section{M-APSK Signal Constellations}\label{sec:apsk}
\begin{figure}[!h]
\begin{center}
\includegraphics[scale=0.7]{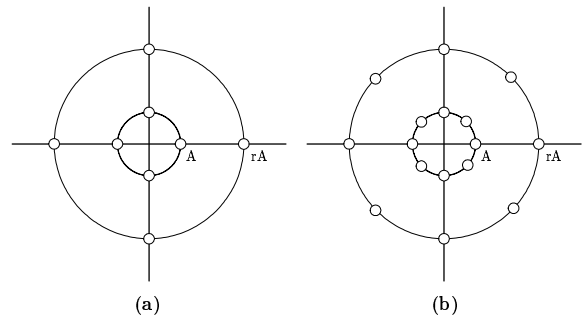}
\end{center}
\vspace{-10pt}\caption{Constellation diagrams with two amplitude values: $A$,$rA$. (a)
$8$-APSK $(2,4)$ and (b) $16$-APSK $(2,8)$.}\label{Fig:figura1}
\end{figure}

We consider APSK constellation diagrams which consist of $N$ different amplitude rings,
each one with $P$ phase values. The amplitude values of the rings differ by a constant
factor $r$ denominated \emph{ring ratio}. Such constellations will be denoted by $M$-APSK
$(N,P)$, with $M=NP$. Fig.\ \ref{Fig:figura1} shows two examples of constellations for
$N=2$ with $P=4$ and $P=8$.


\begin{figure}[!h]
\centerline{\scalebox{.55}{\includegraphics{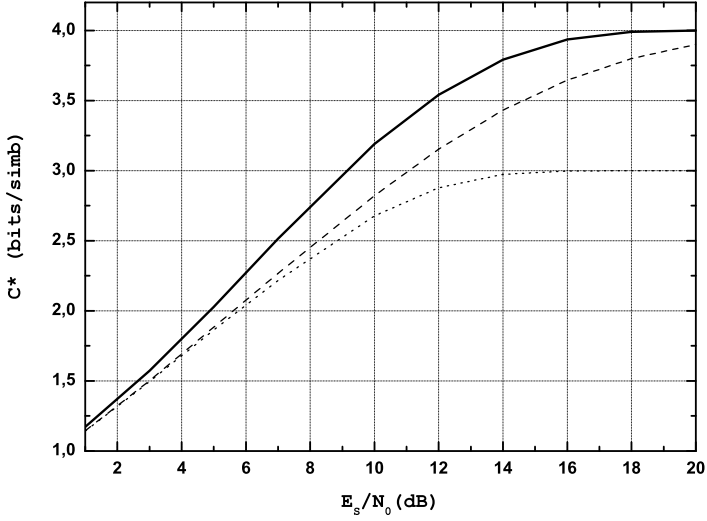}}}
\vspace{-10pt}\caption{Capacities for APSK constellations with optimal ring ratios.
Dotted line:~$8$-APSK$(2,4)$, dashed line:~$16$-APSK$(4,4)$, solid
line:~$16$-APSK$(2,8)$.}\label{Fig:figura2}
\end{figure}

Since it is expected that the noncoherent capacity approaches that of a coherent channel
for large values of block length, we are interested in calculating this capacity. For
fixed SNR, the capacity of an APSK alphabet depends on the ring ratio. For a uniform
input distribution, the capacity of an APSK alphabet, $C^{\ast}$,  can be efficiently
evaluated by Monte Carlo methods \cite{Ungerboeck82}. By doing so, we could obtain the
values of $r$ that maximize $C^{\ast}$ for each SNR. Fig.\ \ref{Fig:figura2} shows
results for three constellations: $8$-APSK$(2,4)$, 16-APSK$(2,8)$ and 16-APSK$(4,4)$.
$E_s$ is the average constellation energy and $N_{0}$ is the one-sided noise spectral
density. The results show that 16-APSK$(2,8)$ has a greater capacity when compared to
16-APSK$(4,4)$.

It was observed that the optimal value of $r$ does not change significantly for SNR's
greater than 2 dB. This observation led us to choose constellations with fixed $r$ in
order to compute bounds on the capacity of noncoherent channels. For $8$-APSK$(2,4)$ and
$16$-APSK$(2,8)$, $r=2,42$ and $r=2,0$ were chosen, respectively. This last value was
also suggested in \cite{Lampe99}.


\section{Channel Model and Definitions}\label{sec:channel_mod}
The input of the channel is a vector of length $L$, ${\mathbf{S}}=\left[
{s_{0},s_{1},...,s_{L-1}}\right],$ whose components $s_l=a_l\exp\left(
{j\phi_{l}}\right)$ represents APSK-modulated symbols. Their average energy is $E_{s}$.
The amplitudes $a_{l}$ can assume one of $N$ possible discrete values and $\phi_{l}$ can
assume one of $P$ discrete phases, so the signal $s_l$ belongs to a $M$-APSK $(N,P)$
constellation. The output is also a vector of length $L,{\mathbf{R}}=\left[
{r_{0},r_{1},...,r_{L-1}}\right],$ whose components may be expressed as
\begin{equation}
r_l=s_l\exp\left({j\theta}\right) + n_l\quad,\quad l=0,1,...,L-1 \label{Eq:saida}
\end{equation}
where $\theta$ is a phase shift introduced by the channel uniformely distributed over the
interval $\left[{\left.{0,2\pi}\right)}\right.$ and $n_{l}$ are independent circularly
symmetric Gaussian noise variables, whose real and imaginary parts are each zero mean
with variance $\sigma^{2}=$ ${N_{0}/2}$. The SNR is then defined as $E_{s}/N_{0}$. In the
following we will use the vectors ${\mathbf{A}}=\left[ {a_{0},a_{1},...,a_{L-1}}\right]
$ and ${\mathbf{\Phi}}=\left[ {\phi_{0},\phi_{1},...,\phi_{L-1}}\right]  $ that can be
defined by using the components $s_{l}$ of $\mathbf{S}$.

Since an input distribution for $\mathbf{S}$ is assumed, we would like to obtain the
\emph{Average Mutual Information} (AMI), $I_{nc}$, of \ the channel described above
using the formula :
\begin{equation}
I_{nc}=I\left(  {\mathbf{S};\mathbf{R}}\right)  =E_{\mathbf{S},\mathbf{R}}%
\log_{2}\left(  {\frac{{P\left(  {\mathbf{R}|\mathbf{S}}\right)  }}{{P\left(
\mathbf{R}\right)  }}}\right)  ,\label{Eq:Inc3}%
\end{equation}
where $E_{\mathbf{S},\mathbf{R}}$ denotes the statistical expectation taken with respect
to variables $\mathbf{S}$ and $\mathbf{R}$. The  transition probability densities
$P(\mathbf{R}|\mathbf{S})$ are given by \cite{Divsalar90} :
\begin{align}
P\left(  {\mathbf{R}|\mathbf{S}}\right)   &  =\frac{1}{{\left(  {2\pi
\sigma^{2}}\right)  ^{L}}}\exp\left[  {\ -\frac{1}{{2\sigma^{2}}}%
\sum\limits_{l=0}^{L-1}{\left(  {\left\vert {r_{l}}\right\vert ^{2}+\left\vert
{s_{l}}\right\vert ^{2}}\right)  }}\right]  \nonumber\\
&  \cdot I_{0}\left(  {\frac{1}{{\sigma^{2}}}\left\vert {\sum\limits_{l=0}%
^{L-1}{r_{l}s_{l}^{\ast}}}\right\vert }\right)  .\label{Eq:pRS}%
\end{align}
where $I_{0}(\cdot)$ is the modified Bessel function of the first kind of order
zero. The probability density $P(\mathbf{R})$ can be obtained by the following
equation:
\begin{equation}
P\left(  \mathbf{R}\right)  =\sum\limits_{\mathbf{S}}{P\left(  {\mathbf{R}%
|\mathbf{S}}\right)  }P\left(  \mathbf{S}\right)  ,\label{Eq:pR}%
\end{equation}
where $P(\mathbf{S})$ is the distribution of the channel input $\mathbf{S}$.

The computing of $I_{nc}$ is rather complicated for large $L$ and $M(=NP)$. \ It is
then appropriate to resort to bounds. As in \cite{Peleg98}, we consider the case
where there exists overlapping of one symbol between consecutive blocks. \ Therefore
the following normalization for the capacity (in bits per modulation symbol) is used
throughout
\begin{equation}
C_{nc}=\frac{{I_{nc}}}{{L-1}}\quad.
\end{equation}

\section{Bounds}\label{sec:bounds}
The steps to derive the bounds are similar to those done in
\cite{Peleg98} for MPSK signals. The phase rotation $\theta$ is viewed as an
additional channel input with AMI $\ I_{v}=I\left(  {\theta,\mathbf{S}%
;\mathbf{R}}\right)  .$ \ Then the chain rule for mutual information \cite{Cover91}
is applied to $I_{v}$ resulting in
\begin{equation}
I_{nc}=I_{v}-I({\mathbf{R};\theta|\mathbf{S)}} \label{Eq:Ivirtual3}%
\end{equation}
and hence,
\begin{equation}
I_{nc}=I\left(  {\mathbf{S};\mathbf{R}|\theta}\right)  -I\left(
{\theta;\mathbf{R}|\mathbf{S}}\right)  +I\left(  {\theta;\mathbf{R}}\right)  .
\label{Eq:Inc4}%
\end{equation}
Equivalent to MPSK signals, the first term is the AMI over the APSK-AWGN
coherent channel while the term $[I\left({\theta;\mathbf{R}|\mathbf{S}%
}\right)-I\left({\theta;\mathbf{R}}\right)]$ represents the degradation due to unknown
$\theta$.

An upper bound on $I_{nc}$ is derived by computing (\ref{Eq:Inc4})
for $\theta$ discretely and uniformly distributed over the same
number of input phases , i.e., $\theta$ has the same distribution
of $\phi_{i}$. Therefore, we have
\begin{equation}
I(\mathbf{S};\mathbf{R}|\theta)\leq\left(  {L-1}\right)  C_{c}\quad,
\label{Eq:LimSup_1termo}%
\end{equation}
where $C_{c}$ is the APSK-AWGN coherent channel capacity. The coefficient $(L-1)$ is
used in (\ref{Eq:LimSup_1termo}) because of overlapping of one symbol in detection.


\begin{figure}[!h]
\centerline{\scalebox{.4}{\includegraphics{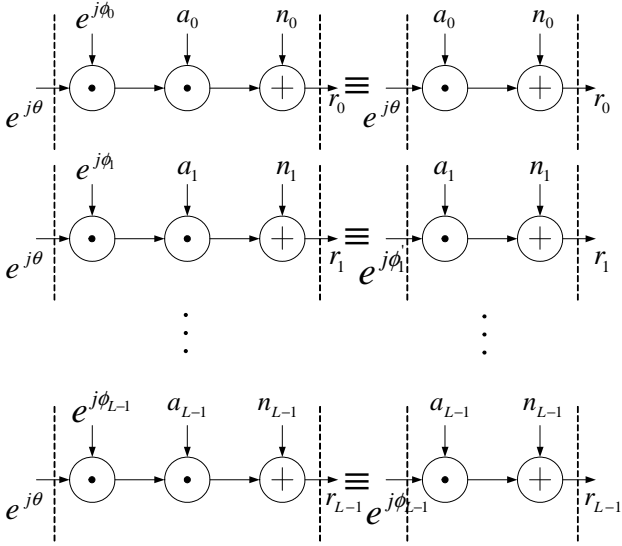}}} \vspace{-10pt}\caption{Channel
model for evaluating $\ I(\theta;\mathbf{R})$.}\label{Fig:figura3}
\end{figure}

For evaluating $\ I(\theta;\mathbf{R})$, we will consider the channel model shown in
Fig.\ \ref{Fig:figura3}. This is a \emph{Single Input Multiple Output} (SIMO) channel
\cite{Tse04}, with $\theta$ as the single input. Define
$\phi_{l}^{^{\prime}}=(\theta\oplus\phi_{l}),\quad l=1,2,...,L-1$, where $\oplus$ is sum
modulus $2\pi$ . Since the $\phi_{l}$ are u.i.i.d. variables, the $\phi _{l}^{^{\prime}}$
are independent of $\theta$ implying that $p(r_{l}|\theta),$ $l=1,2,...,L-1$, are also
independent of $\theta$. Consequently, we have
\[
I\left(  {\theta;r_{l}}\right)  =0,\quad l=1,2,...,L-1\quad
,\label{Eq:LimSup_3termo_1}%
\]
and, therefore, only the $r_{0}$ coordinate carries information on $\theta$. Then,
\begin{equation}
I\left(  {\theta;\mathbf{R}}\right)  =I\left(  {\theta;r_{0}}\right)  .
\label{Eq:LimSup_3termo_3}%
\end{equation}
The APSK reference symbol $s_{0}$ $=a_{0}\exp\left(  {j\phi}_{0}\right)  $ can
assume any of the $M(=NP)$ values. Accordingly, we write
\[
I\left(  {\theta;r_{0}}\right)  =I\left(
{\theta;r_{0}|s_{0}}\right) =
{\displaystyle\sum\limits_{k=0}^{M-1}}
{P_{s_{0}}}\left(  k\right)  I\left(  {\theta;r_{0}\left\vert {s_{0}%
=k}\right.  }\right)  \label{Eq:LimSup_3termo_4}
\]
or
\begin{equation}
I\left(  {\theta;r_{0}}\right)  =
{\displaystyle\sum\limits_{k=0}^{N-1}}
{P_{a_{0}}}\left(  k\right)  I\left(  {\theta;r_{0}\left\vert {a_{0}%
=k}\right.  }\right)~, \label{Eq:LimSup_3termo_5}
\end{equation}
due to the fact that phase rotations do not change the mutual information.

From (\ref{Eq:LimSup_3termo_5}), we conclude that $I(\theta;\mathbf{R})$ is
calculated as an average of capacities of PSK modulations over a coherent AWGN
channel. For example, considering 8-APSK$(2,4)$ we have
\[
I\left(  {\theta;\mathbf{R}}\right)  =\frac{1}{2}~C_{c-4PSK(A)}+\frac{1}
{2}~C_{c-4PSK(rA)}~,\label{Eq:LimSup_3termo_6}
\]
where $C_{c-4PSK(A)}\ $and $C_{c-4PSK(rA)}$ are the 4-PSK channel capacities for two
amplitudes, $A$ and $rA$, respectively.

Finally, $I({\theta;\mathbf{R}|\mathbf{S}})$ is given by the following equation
\cite{Blahut87}:

\begin{equation}
I\left(  {\theta;\mathbf{R}|\mathbf{S}}\right)  =
{\displaystyle\sum\limits_{\bf{\alpha}} {P_{\mathbf{S}}} \left(
\bf{\alpha}\right) I\left( {\theta;\mathbf{R}|\mathbf{S}%
=\bf{\alpha}}\right) }~. \label{Eq:LimSup_2}
\end{equation}

By using again the concept of a SIMO channel,
$I({\theta;\mathbf{R}|\mathbf{S})}$ is obtained by computing
$I({\theta;r_{l}|s_{l})}$, the AMI of the $l$-th component, with
SNR increased by a factor of $L$.   As above, we have

\begin{equation}
I\left(  {\theta;r_{l} |s_{l} } \right)  = I\left(  {\theta;r_{l} |a_{l} } \right) =
\sum\limits_{k} {P_{a_{l} } \left(  k \right)  } I\left( {\theta;r_{l} |a_{l} = k}
\right)  . \label{Eq:LimSup_23}
\end{equation}

Therefore, (\ref{Eq:LimSup_23}) is evaluated using the same reasoning that was applied to
computation of (\ref{Eq:LimSup_3termo_5}).

The lower bound is also obtained starting with (\ref{Eq:Inc4}), but knowing that the
unknown phase $\theta$ is a continuous uniformly distributed variable. Following
\cite{Peleg98}, we incorporate the inequality
\[
I\left(  {\theta;\mathbf{R}} \right)  \ge I\left(  {\theta;r_{0} } \right)
\label{Eq:LimSup_3termo_2}%
\]
to (\ref{Eq:Inc4}) yielding:
\begin{equation}
I_{nc} \ge I\left(  {\mathbf{S};\mathbf{R}|\theta} \right)  - I\left(
{\theta;\mathbf{R}|\mathbf{S}} \right)  + I\left(  {\theta;r_{0}} \right)  .
\label{Eq:LimInf}
\end{equation}

The first term of the right hand side of (\ref{Eq:LimInf}) is identical to the first term
of the upper bound. The third term, $I({\theta;r_{0}})$, is also given by
(\ref{Eq:LimSup_3termo_5}) but with a continuous $\theta$. Each AMI
$I({\theta;r_{0}|a_{0}=k)}$ equals the capacity of a coherent continuous input phase
modulated channel \cite{Wyner66}.  The second term of (\ref{Eq:LimInf}) is also
equivalent to the second term of the upper bound, except that we have to calculate
capacities for a channel with a single continuous input. All these capacities were
evaluated efficiently using Monte Carlo methods.

\section{Numerical Results}\label{sec:conclusion}

\begin{figure}[!h]
\centerline{\scalebox{.55}{\includegraphics{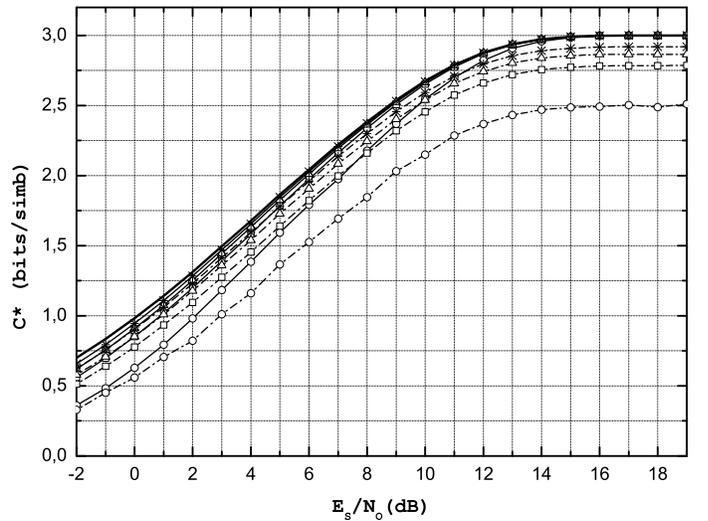}}} \vspace{-10pt}\caption{Bounds
on the capacity of the noncoherent $8$-APSK$(2,4)$-AWGN channel. $\circ$~: $L=2$,
$\square$~: $L=8$, $\vartriangle$~: $L=16$, $\ast$~: $L=32$, \textbf{---}~:
$8$-APSK$(2,4)$-AWGN coherent channel capacity.}\label{Fig:figura5}
\end{figure}

\begin{figure}[!h]
\centerline{\scalebox{.55}{\includegraphics{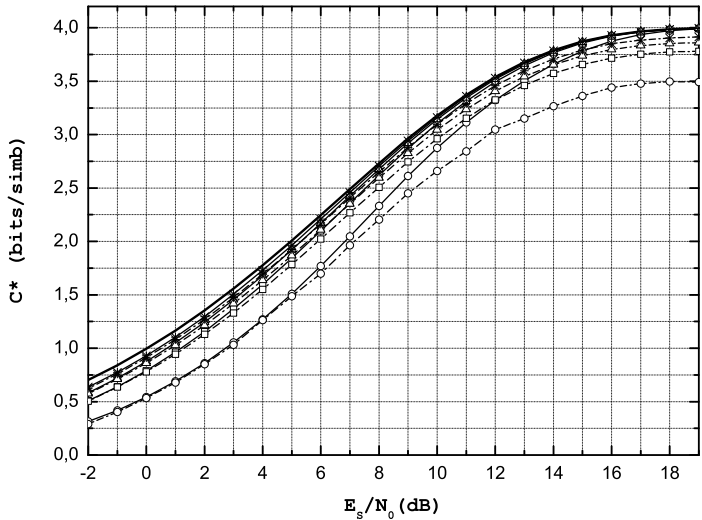}}} \vspace{-10pt}\caption{Bounds
on the capacity of the noncoherent $16$-APSK$(2,8)$-AWGN channel. $\circ$~: $L=2$,
$\square$~: $L=8$, $\vartriangle$~: $L=16$, $\ast$~: $L=32$, \textbf{---}~:
$16$-APSK$(2,8)$-AWGN coherent channel capacity.}\label{Fig:figura6}
\end{figure}

 Figs.\ \ref{Fig:figura5} and \ref{Fig:figura6} illustrate the results for
$8$-APSK$(2,4)$ and $16$-APSK$(2,8)$ constellations, respectively. Solid lines represent
results for the upper bounds while dashed lines represent them for the lower bounds. For
$8$-APSK$(2,4)$ and $L=2$, the bounds come close together with SNR's less than $0$ dB
whereas for $16$-APSK$(2,8)$ and $L=2$ this happens with SNR's less than $6$ dB. It can
be seen that as $L$ increases, the bounds become close to coherent channel capacity.
Moreover, for $L=8,16,32$, the bounds come close together over a broad range of SNR's
(the difference between the upper and lower bounds is less than $0.1$ bit/symbol).
Therefore, we can conclude that the coherent capacity is approached.

\vspace{-1cm}


\section*{Acknowledgment}

The authors would like to thank M. Peleg for helping us to
understand one of the results in \cite{Peleg98}. This work was
supported in part by The State of Sao Paulo Research Foundation
(FAPESP) under Grant 03/05385-6.



\begin{thebibliography}{1}

\bibitem{Divsalar90}
D.~Divsalar and M.K.~Simon, ``Multiple-symbol differential
detection of MPSK,''  {\em IEEE Trans. Commun.}, vol. 38, pp.
300-308, Mar. 1990.

\bibitem{Cahn60}
C.~Cahn,
\newblock {``Combined Digital Phase and Amplitude Modulation Communication Systems,''}
\newblock in {\em IRE Trans. on Communications Systems, vol. 8, pp. 150-155, Sept. 1960.}

\bibitem{Lampe99}
L.~Lampe and R.~Fischer,
\newblock {``Comparison and optimization of differentially encoded transmission on fading channels,''}
\newblock in {Proceedings of {\em $3^{rd}$ International
Symposium on Power-Line Communications and its Applications
(ISPLC'99)}, pp. 107-113, Lancaster, UK, Mar. 1999.}

\bibitem{Peleg98}
M.~Peleg and S.~Shamai (Shitz), ``On the capacity of the blockwise
incoherent MPSK channel,'' in {\em IEEE Trans. Commun.}, vol. 46,
pp. 603-609, May 1998.

\bibitem{Ungerboeck82}
G.~Ungerboeck,
\newblock {``Channel coding with multilevel/phase signals,''}
\newblock in {{\em IEEE Trans. Inform. Theory}, vol. IT-28, pp. 55-67, Jan. 1982.}

\bibitem{Cover91}
T.~Cover,
\newblock {Elements of Information Theory.}
\newblock {New York : Wiley, 1991.}

\bibitem{Tse04}
D.~Tse and P.~Viswanath,
\newblock {\emph{Fundamentals of Wireless Communications.}}
\newblock {University of California, Berkeley, 2004.
available at http://www-inst.eecs.berkeley.edu/$\sim$ee224b/sp04/,
Access: 01 Oct. 2004.}

\bibitem{Blahut87}
R.Blahut,
\newblock {Principles and Practice of Information Theory.}
\newblock {New York : Addison-Wesley, 1987.}

\bibitem{Wyner66}
A.~D.~Wyner,
\newblock {``Bounds on communication with polyphase coding,''}
\newblock {\em Bell Syst. Tech. J., vol. XLV, pp. 523-559, Apr. 1966.}

\end{thebibliography}
\end{document}